\documentclass[11pt]{article}
\usepackage[a4paper,margin=0.85in]{geometry}
\usepackage{amsmath,amssymb,amsthm}
\usepackage{dsfont}
\usepackage{booktabs}
\usepackage{microtype}
\usepackage{titlesec}
\usepackage{placeins}
\usepackage{cite}
\usepackage[colorlinks=true,linkcolor=blue,citecolor=blue,urlcolor=blue]{hyperref}
\titlespacing*{\paragraph}{0pt}{1.2ex plus .5ex minus .2ex}{1em}
\newcommand{\ket}[1]{\lvert #1\rangle}
\newcommand{\bra}[1]{\langle #1\rvert}
\newcommand{\braket}[2]{\langle #1|#2\rangle}
\newcommand{\proj}[1]{\lvert #1\rangle\!\langle #1\rvert}
\newcommand{\id}{\mathds{1}}
\newcommand{\cX}{\mathcal X}
\newcommand{\cY}{\mathcal Y}
\newtheorem{theorem}{Theorem}
\newtheorem{proposition}{Proposition}
\title{\Large\bfseries\boldmath Perfect discrimination of an orthogonal $2\otimes2$ basis may require collective separable measurements on two copies}
\author{Somshubhro Bandyopadhyay\thanks{\texttt{som@jcbose.ac.in}}\\[4pt]
\small\itshape Department of Physical Sciences, Bose Institute, Kolkata 700091, India}
\date{}

\begin{document}
\maketitle

\begin{abstract}
We consider an orthogonal basis of $\mathbb C^2\otimes\mathbb C^2$ with three entangled states and one product state. This basis is not perfectly distinguishable by LOCC or even by separable measurements when a single copy of the unknown state is available. We show that with two copies no adaptive separable protocol, which measures the copies one at a time, distinguishes the basis perfectly. We then show that a collective separable measurement on the two copies does. Whether two copies suffice for LOCC remains open.
\end{abstract}

\section{Introduction}

One of the central problems in quantum information theory is quantum state discrimination by local operations and classical communication (LOCC)~\cite{Chitambar2014}. In the simplest case of the problem, we suppose that two distant observers Alice and Bob share a pure state drawn from a set of mutually orthogonal states. While they have complete knowledge of the set, they do not know which particular state they hold. Their objective is to determine the state by LOCC. That is, they may perform any operation on the subsystems in their possession and communicate classically, but they cannot exchange quantum systems. For some sets they can accomplish this task with certainty~\cite{Walgate2000}, whereas for others they cannot~\cite{BennettNWE,Ghosh2001,WalgateHardy,Horodecki2003,GhoshMES2004,Fan2004,Watrous2005,YuDuanYing2012}, even though the states are perfectly distinguishable by a global measurement. Such sets are said to be nonlocal. If all the states of such a set are product states, the phenomenon is called nonlocality without entanglement~\cite{BennettNWE,BennettUPB,DiVincenzo2003,NisetCerf2006,Halder2019}.

One of the fundamental results is that any two orthogonal pure states can be perfectly distinguished by LOCC~\cite{Walgate2000}. This holds irrespective of the entanglement of the states, the local dimensions and the number of parties. However, sets containing three or more states are not always locally distinguishable~\cite{Ghosh2001,Nathanson2005,BandyopadhyayWalgate2009,YuDuanYing2012,CosentinoRusso2014}. In particular, an orthogonal basis of a finite-dimensional bipartite system that contains at least one entangled state cannot be perfectly distinguished by LOCC~\cite{Horodecki2003}.

In the standard formulation of the problem, a single copy of the unknown state is available. If we relax the single-copy condition, then with sufficiently many copies any locally indistinguishable set becomes perfectly distinguishable by LOCC. From the two-state result it follows that at most $(N-1)$ copies are required to perfectly distinguish any set of $N$ orthogonal pure states by LOCC~\cite{Walgate2000,Bandyopadhyay2011}, as each copy can be used to rule out at least one state. On the other hand, many well-known locally indistinguishable sets can be perfectly distinguished by LOCC with just two copies. These include the Bell basis~\cite{Ghosh2001}, the canonical nonmaximally entangled basis in $\mathbb C^2\otimes\mathbb C^2$~\cite{Banik2021}, and the canonical maximally entangled basis in $\mathbb C^d\otimes\mathbb C^d$~\cite{GhoshMES2004}.

Thus a basic question, first raised in Ref.~\cite{BennettNWE}, is whether two copies are always sufficient for perfect LOCC discrimination of orthogonal states, or whether some set requires three or more. For multipartite systems, it was recently shown that there exist bases for which the required number of copies grows at least as the square root of the number of parties, even for collective separable measurements~\cite{Li2026}. For bipartite systems, however, the question remains open even for separable and positive-partial-transpose (PPT) measurements~\cite{BCJRWY2015}. Since two copies are always sufficient for distinguishing three states, the first nontrivial case is that of four states. In this paper we address it for an orthogonal basis of $\mathbb C^2\otimes\mathbb C^2$.

Consider therefore the problem of distinguishing the states of an orthogonal basis of $\mathbb C^2\otimes\mathbb C^2$ by LOCC. We suppose that the basis is locally indistinguishable and that two copies of the unknown state are now available. The copies may be used in two ways. In the adaptive approach~\cite{Banik2021}, they are used one at a time, where the measurement on the second copy may depend on the outcome for the first. Ref.~\cite{Banik2021} identified two-qubit bases that cannot be perfectly distinguished in this way. In the collective approach all possible two-qubit measurements on the local qubits may be performed. This therefore contains the adaptive approach.

Though intuitively LOCC is easy to understand, its mathematical structure is difficult to characterize. Therefore, it is sometimes helpful to consider the larger and more tractable class of separable (SEP) measurements~\cite{Duan2009,DuanXinYing2010,BandyopadhyayNathanson2013,ChitambarDuanHsieh2014,BCJRWY2015,Cohen2015,Yang2015}, as we do in this paper. Since LOCC is a subset of SEP, if separable measurements fail to perfectly distinguish a set of states, then the same holds for LOCC. But the inclusion is strict~\cite{BennettNWE,BennettUPB,DiVincenzo2003,NisetCerf2006,Halder2019}. A separable measurement can perfectly distinguish the states, but there need not be an LOCC protocol that does the same.

The basis that we consider contains three entangled states and one product state. With one copy it cannot be perfectly distinguished by separable measurements~\cite{Duan2009}, and hence not by LOCC. We first show that with two copies every adaptive SEP protocol fails to distinguish the states perfectly. We then show that a collective separable measurement on two copies succeeds. Whether two copies suffice for LOCC remains open.

\section{Separable measurements}\label{sec:sepmeas}
Let $\mathcal H_A$ and $\mathcal H_B$ be finite-dimensional Hilbert spaces. A measurement $\{E_i\}$ on $\mathcal H_A\otimes\mathcal H_B$ is separable~\cite{BennettNWE,Watrous2005} if
\begin{equation}\label{eq:sepdef}
E_i=\sum_jA_{ij}\otimes B_{ij},
\end{equation}
where the operators $A_{ij}$ and $B_{ij}$ are positive semidefinite.
Since every positive semidefinite operator is a sum of rank-one projectors with positive weights, $A_{ij}$ and $B_{ij}$ may be taken to be rank one. Each term $A_{ij}\otimes B_{ij}$ is implemented by the Kraus operator $\sqrt{A_{ij}}\otimes\sqrt{B_{ij}}$, so a separable measurement is a separable operation with a classical outcome~\cite{Duan2009}.

A separable measurement $\{E_i\}$ perfectly distinguishes states $\ket{\phi_1},\dots,\ket{\phi_N}$ if the index set can be partitioned into sets $\mathcal S_1,\dots,\mathcal S_N$ such that
\begin{equation}\label{eq:sepdist}
\bra{\phi_k}\Bigl(\sum_{i\in\mathcal S_l}E_i\Bigr)\ket{\phi_k}=\delta_{kl}\qquad(k,l=1,\dots,N).
\end{equation}

Suppose now that $\mathcal H_A=\mathcal H_{A_1}\otimes\mathcal H_{A_2}$ and $\mathcal H_B=\mathcal H_{B_1}\otimes\mathcal H_{B_2}$. An adaptive SEP protocol is a separable measurement $\{M_l\}$ on $\mathcal H_{A_1}\otimes\mathcal H_{B_1}$ followed, for each outcome $l$, by a separable measurement $\{N_{m|l}\}$ on $\mathcal H_{A_2}\otimes\mathcal H_{B_2}$. Its elements are
\begin{equation}\label{eq:adaptive}
E_{lm}=M_l^{(1)}\otimes N_{m|l}^{(2)},
\end{equation}
where the superscripts label the copies. A separable measurement on $\mathcal H_A\otimes\mathcal H_B$ without this restriction is the most general form of SEP measurement and in this paper we will sometimes refer to it as collective.

\section{The basis}\label{sec:basis}
Consider the orthonormal basis\footnote{This basis was suggested by John Watrous in 2014 as a possible example for which two copies may not be sufficient for perfect local discrimination.} of $\mathbb C^2\otimes\mathbb C^2$
\begin{equation}\label{eq:omega-basis}
\ket{\psi_k}=\frac1{\sqrt3}\bigl(\ket{00}+\omega^k\ket{01}+\omega^{2k}\ket{10}\bigr),\quad k=0,1,2,\qquad \ket{\psi_3}=\ket{11},
\end{equation}
where $\omega=e^{2\pi i/3}$.
The first three states are equally entangled, with concurrence $2/3$~\cite{Wootters1998}. With one copy this basis cannot be perfectly distinguished by separable measurements~\cite{Duan2009}, and hence not by LOCC~\cite{WalgateHardy,Horodecki2003}.

With two copies the states are
\begin{equation}\label{eq:twocopy}
\ket{\Psi_k}=\ket{\psi_k}\otimes\ket{\psi_k}\in(\cX_1\otimes\cY_1)\otimes(\cX_2\otimes\cY_2),\qquad k=0,\dots,3,
\end{equation}
with $\cX_i=\cY_i=\mathbb C^2$.

We now write the states~\eqref{eq:twocopy} as vectors in $\cX\otimes\cY$, where $\cX=\cX_1\otimes\cX_2$ and $\cY=\cY_1\otimes\cY_2$, as
\begin{subequations}\label{eq:PsiXY}
\begin{align}
\ket{\Psi_k}={}&\tfrac13\Bigl[\bigl(\ket{00}\ket{00}+\ket{01}\ket{10}+\ket{10}\ket{01}\bigr)+\omega^k\bigl(\ket{00}\ket{01}+\ket{00}\ket{10}+\ket{11}\ket{00}\bigr)\nonumber\\
&+\omega^{2k}\bigl(\ket{01}\ket{00}+\ket{10}\ket{00}+\ket{00}\ket{11}\bigr)\Bigr]\qquad(k=0,1,2),\\
\ket{\Psi_3}={}&\ket{11}\ket{11}.
\end{align}
\end{subequations}

The states $\ket{\psi_0},\ket{\psi_1},\ket{\psi_2}$ are related by local unitaries as follows. Let
\begin{equation}\label{eq:UAUB}
U_A=\begin{pmatrix}1&0\\0&\omega^2\end{pmatrix},\qquad U_B=\begin{pmatrix}1&0\\0&\omega\end{pmatrix}.
\end{equation}
Then
\begin{subequations}\label{eq:Ustates}
\begin{align}
(U_A\otimes U_B)\ket{\psi_k}&=\ket{\psi_{k\oplus1}}\qquad(k=0,1,2),\\
(U_A\otimes U_B)\ket{\psi_3}&=\ket{\psi_3},
\end{align}
\end{subequations}
where $\oplus$ denotes addition mod 3 on $\{0,1,2\}$.

For two copies let
\begin{equation}\label{eq:W}
W=(U_A\otimes U_A)\otimes(U_B\otimes U_B)
\end{equation}
act on $\cX\otimes\cY$. Then
\begin{subequations}\label{eq:Wstates}
\begin{align}
W\ket{\Psi_k}&=\ket{\Psi_{k\oplus1}}\qquad(k=0,1,2),\\
W\ket{\Psi_3}&=\ket{\Psi_3}.
\end{align}
\end{subequations}
This symmetry, as we will see, is what makes the problem tractable.

\section{Results}\label{sec:results}
Ref.~\cite{Banik2021} states that no orthogonal basis of $\mathbb C^2\otimes\mathbb C^2$ containing exactly three entangled states can be perfectly distinguished by adaptive LOCC with two copies. Since LOCC is a strict subset of SEP, the same may not hold for adaptive SEP protocols. Moreover, the basis~\eqref{eq:omega-basis} is not a member of the family of bases for which Ref.~\cite{Banik2021} analyzes adaptive SEP protocols, even up to local unitaries.\footnote{In that family three basis states cannot have equal nonzero concurrence while the fourth is a product state.}

Our first result is that, with two copies, the basis~\eqref{eq:omega-basis} is not perfectly distinguishable by an adaptive SEP protocol.

\begin{proposition}\label{prop:adaptive}
The states $\ket{\Psi_0},\dots,\ket{\Psi_3}$ cannot be perfectly distinguished by an adaptive SEP protocol.
\end{proposition}

The proof is given in Sec.~\ref{sec:adaptive}.

A general separable measurement on $\cX\otimes\cY$ has elements $\sum_jA_{ij}\otimes B_{ij}$ with $A_{ij}$ acting on $\cX_1\otimes\cX_2$ and $B_{ij}$ on $\cY_1\otimes\cY_2$. Our main result is that the states are perfectly distinguishable by such measurements.

\begin{theorem}\label{thm:sep}
The states~\eqref{eq:twocopy} are perfectly distinguishable by a separable measurement on $\cX\otimes\cY$.
\end{theorem}

The proof of Theorem~\ref{thm:sep} is constructive.

\begin{proposition}[Separable measurement]\label{prop:collective}
Consider the separable measurement $\{P_0,\dots,P_4\}$ on $\cX\otimes\cY$ with
\begin{align}
&P_0=\sum_{r=1}^{5}w_r\,\Pi_r,\qquad P_1=WP_0W^\dagger,\qquad P_2=W^\dagger P_0W,\qquad P_3=\proj{11}\otimes\proj{11},\nonumber\\
&P_4=\tfrac34\left(\proj{11}\otimes\proj{\Psi^+}+\proj{\Psi^+}\otimes\proj{11}\right)+\tfrac14\left(M\otimes\proj{\Psi^-}+\proj{\Psi^-}\otimes M\right),
\end{align}
where
\begin{equation}\label{eq:PiM}
\Pi_r=\sum_{i=1}^4\proj{x_{ri}y_{ri}},\qquad M=\proj{00}+\proj{11}+2\,\proj{\Psi^+},
\end{equation}
with $\ket{\Psi^\pm}=\tfrac1{\sqrt2}(\ket{01}\pm\ket{10})$ and the normalized vectors $\ket{x_{ri}}\ket{y_{ri}}$ and weights $w_r$ of Table~\ref{tab:reps}. Then
\begin{equation}
P_i\ket{\Psi_j}=\delta_{ij}\ket{\Psi_j},\qquad P_4\ket{\Psi_j}=0\qquad(i,j=0,\dots,3).
\end{equation}
\end{proposition}

\begin{table}[t]
\centering
\footnotesize
\renewcommand{\arraystretch}{1.3}
\begin{tabular}{c c c l l}
\toprule
$r$ & $w_r$ & $i$ & $\ket{x_{ri}}\in\cX=\cX_1\otimes\cX_2$ & $\ket{y_{ri}}\in\cY=\cY_1\otimes\cY_2$\\
\midrule
1 & $\frac{1}{6}$ & 1 & $\frac1{\sqrt{3}}\bigl(\sqrt2\,\ket{\Phi^+}+\ket{10}\bigr)$ & $\ket{00}$\\
 & & 2 & $\ket{00}$ & $\frac1{\sqrt{3}}\bigl(\sqrt2\,\ket{\Phi^+}+\ket{10}\bigr)$\\
 & & 3 & $\frac1{\sqrt{3}}\bigl(\sqrt2\,\ket{\Phi^+}+\ket{01}\bigr)$ & $\ket{00}$\\
 & & 4 & $\ket{00}$ & $\frac1{\sqrt{3}}\bigl(\sqrt2\,\ket{\Phi^+}+\ket{01}\bigr)$\\
\midrule
2 & $\frac{1}{4}$ & 1 & $\frac1{\sqrt{3}}\bigl(\ket{00}+\sqrt2\,\ket{\Psi^-}\bigr)$ & $\ket{1+}$\\
 & & 2 & $\ket{1+}$ & $\frac1{\sqrt{3}}\bigl(\ket{00}+\sqrt2\,\ket{\Psi^-}\bigr)$\\
 & & 3 & $\frac1{\sqrt{3}}\bigl(\ket{00}-\sqrt2\,\ket{\Psi^-}\bigr)$ & $\ket{+1}$\\
 & & 4 & $\ket{+1}$ & $\frac1{\sqrt{3}}\bigl(\ket{00}-\sqrt2\,\ket{\Psi^-}\bigr)$\\
\midrule
3 & $\frac{1}{6}$ & 1 & $\frac1{\sqrt{3}}\bigl(\ket{00}+\ket{01}-\omega\ket{10}\bigr)$ & $\ket{\chi_1\chi_1}$\\
 & & 2 & $\ket{\chi_1\chi_1}$ & $\frac1{\sqrt{3}}\bigl(\ket{00}+\ket{01}-\omega\ket{10}\bigr)$\\
 & & 3 & $\frac1{\sqrt{3}}\bigl(\ket{00}-\omega\ket{01}+\ket{10}\bigr)$ & $\ket{\chi_1\chi_1}$\\
 & & 4 & $\ket{\chi_1\chi_1}$ & $\frac1{\sqrt{3}}\bigl(\ket{00}-\omega\ket{01}+\ket{10}\bigr)$\\
\midrule
4 & $\frac{1}{6}$ & 1 & $\frac1{\sqrt{3}}\bigl(\ket{00}+\ket{01}-\omega^2\ket{10}\bigr)$ & $\ket{\chi_2\chi_2}$\\
 & & 2 & $\ket{\chi_2\chi_2}$ & $\frac1{\sqrt{3}}\bigl(\ket{00}+\ket{01}-\omega^2\ket{10}\bigr)$\\
 & & 3 & $\frac1{\sqrt{3}}\bigl(\ket{00}-\omega^2\ket{01}+\ket{10}\bigr)$ & $\ket{\chi_2\chi_2}$\\
 & & 4 & $\ket{\chi_2\chi_2}$ & $\frac1{\sqrt{3}}\bigl(\ket{00}-\omega^2\ket{01}+\ket{10}\bigr)$\\
\midrule
5 & $\frac{5}{24}$ & 1 & $\frac1{\sqrt{15}}\bigl(3\ket{00}-\sqrt2\,\omega(1-\omega)\ket{\Psi^-}\bigr)$ & $\frac1{\sqrt{3}}\bigl(\ket{01}+\omega\ket{10}-\omega^2\ket{11}\bigr)$\\
 & & 2 & $\frac1{\sqrt{3}}\bigl(\ket{01}+\omega\ket{10}-\omega^2\ket{11}\bigr)$ & $\frac1{\sqrt{15}}\bigl(3\ket{00}-\sqrt2\,\omega(1-\omega)\ket{\Psi^-}\bigr)$\\
 & & 3 & $\frac1{\sqrt{15}}\bigl(3\ket{00}+\sqrt2\,\omega(1-\omega)\ket{\Psi^-}\bigr)$ & $\frac1{\sqrt{3}}\bigl(\ket{01}+\omega^2\ket{10}-\omega\ket{11}\bigr)$\\
 & & 4 & $\frac1{\sqrt{3}}\bigl(\ket{01}+\omega^2\ket{10}-\omega\ket{11}\bigr)$ & $\frac1{\sqrt{15}}\bigl(3\ket{00}+\sqrt2\,\omega(1-\omega)\ket{\Psi^-}\bigr)$\\
\bottomrule
\end{tabular}
\caption{The $20$ product vectors $\ket{x_{ri}}\ket{y_{ri}}$ of $P_0$, normalized, in five groups. Every vector of group $r$ carries the weight $w_r$. Here $\ket\pm=\tfrac1{\sqrt2}(\ket0\pm\ket1)$, $\ket{\chi_k}=\tfrac1{\sqrt2}(\ket0-\omega^k\ket1)$ $(k=1,2)$, $\ket{\Phi^\pm}=\tfrac1{\sqrt2}(\ket{00}\pm\ket{11})$ and $\ket{\Psi^\pm}=\tfrac1{\sqrt2}(\ket{01}\pm\ket{10})$.}
\label{tab:reps}
\end{table}

The outcome $4$ never occurs, and $P_4$ only completes the measurement. Proposition~\ref{prop:collective} therefore proves Theorem~\ref{thm:sep}. Proposition~\ref{prop:collective} is proved in Sec.~\ref{sec:sep}.

\section{Proof of Proposition~\ref{prop:adaptive}}\label{sec:adaptive}

We now prove Proposition~\ref{prop:adaptive}.\footnote{Our argument makes no assumption about which states each outcome of the first measurement rules out. The adaptive SEP analysis of Ref.~\cite{Banik2021} considers partitions of the basis into one-versus-three or two-versus-two groups, as well as outcomes that eliminate no state.} We first show that every separable measurement on a copy has an outcome that eliminates at most one of the four basis states. The proof is completed by showing that after such an outcome the remaining states cannot be perfectly distinguished by a separable measurement on the second copy.

Let $\{M_l\}$ be such a measurement. Each $M_l$ can be written as a sum of rank-one product projectors with positive weights (Sec.~\ref{sec:sepmeas}),
\begin{equation}\label{eq:sep-complete}
M_l=\sum_mw_{lm}\proj{x_{lm}y_{lm}},\qquad w_{lm}>0 .
\end{equation}
The completeness condition is given by
\begin{equation}\label{eq:completeness1}
\sum_{l,m}w_{lm}\proj{x_{lm}y_{lm}}=\id .
\end{equation}
Suppose every outcome eliminates at least two basis states. If the outcome $l$ eliminates two states, then every $\ket{x_{lm}y_{lm}}$ is orthogonal to them. Therefore, each $\ket{x_{lm}y_{lm}}$ lies in the span of the other two basis states.

Now, a vector $\alpha\ket{\psi_i}+\beta\ket{\psi_j}=\sum_{a,b=0}^{1}v_{ab}\ket{ab}$ is a product vector if and only if
\[
v_{00}v_{11}-v_{01}v_{10}=0.
\]
Then a straightforward calculation shows that each of the six spans contains exactly two product vectors (up to normalization).\footnote{For example, $\mathrm{span}\{\ket{\psi_0},\ket{\psi_3}\}$ contains $\ket{11}$ and $(\ket0+\ket1)(\ket0+\ket1)$, and $\mathrm{span}\{\ket{\psi_1},\ket{\psi_2}\}$ contains $\ket0(\ket0-\ket1)$ and $(\ket0-\ket1)\ket0$. Since $U_A\otimes U_B$ maps $\ket{\psi_k}$ to $\ket{\psi_{k\oplus1}}$ and leaves $\ket{\psi_3}$ unchanged by Eq.~\eqref{eq:Ustates}, applying it once and twice to these four vectors gives the product vectors in the remaining four spans.} Since $\ket{\psi_3}=\ket{11}$ lies in three of the spans, there are ten distinct product vectors altogether,
\begin{equation}\label{eq:prodvecs}
\ket{11},\qquad \ket0\bigl(\ket0+c\ket1\bigr),\qquad \bigl(\ket0+c\ket1\bigr)\ket0,\qquad \bigl(\ket0+a\ket1\bigr)\bigl(\ket0+b\ket1\bigr),
\end{equation}
with $c\in\{-1,-\omega,-\omega^2\}$ and $(a,b)\in\{(1,1),(\omega^2,\omega),(\omega,\omega^2)\}$. Every $\ket{x_{lm}y_{lm}}$ is therefore one of the vectors in \eqref{eq:prodvecs} (up to normalization). The completeness condition~\eqref{eq:completeness1} must then hold with vectors from this set alone, with appropriate weights. In particular, the set $\{\ket{x_{lm}y_{lm}}\}$ may consist of all the vectors in \eqref{eq:prodvecs} or a proper subset.

We now show that the completeness condition~\eqref{eq:completeness1} cannot be satisfied with vectors from \eqref{eq:prodvecs}. Using Eq.~\eqref{eq:completeness1}, for any operator $H$ it holds that
\begin{equation}
\operatorname{Tr}H=\sum_{l,m}w_{lm}\bra{x_{lm}y_{lm}}H\ket{x_{lm}y_{lm}} .
\end{equation}
If we can find an operator $H$ with nonzero trace but zero expectation value on every vector in \eqref{eq:prodvecs}, then the completeness condition cannot be satisfied.\footnote{A similar use of the completeness condition appears in Ref.~\cite{Watrous2005}.} Such operators exist and one example is (Appendix~\ref{app:witness})
\begin{equation}\label{eq:H}
H=2\proj{00}-2\proj{01}-2\proj{10}+\ket{00}\!\bra{11}+\ket{11}\!\bra{00}.
\end{equation}
Observe that $\operatorname{Tr}H=-2$. Now it is easy to check that $\bra vH\ket v=0$ for every vector $\ket v$ in \eqref{eq:prodvecs}. Therefore, there does not exist a separable measurement such that every outcome eliminates at least two basis states. This implies that every separable measurement has at least one outcome that eliminates at most one of the four basis states.

Now whenever such an outcome is obtained on the first copy, a separable measurement on the second copy must perfectly distinguish at least three basis states. Every outcome of this measurement must therefore eliminate at least two of the original four basis states. But we have just shown that no separable measurement on a single copy can have this property.
This proves Proposition~\ref{prop:adaptive}.

\section{Proof of Proposition~\ref{prop:collective}}\label{sec:sep}
We prove Proposition~\ref{prop:collective}, from which Theorem~\ref{thm:sep} follows. The proof is a construction in three steps. The symmetry of the states reduces the problem to finding a single operator. We then translate the conditions on this operator into conditions on product vectors and on matrix elements. Finally we find vectors and weights that satisfy them.

\paragraph{Reduction to a single operator.}
The objective is to find a separable measurement $\{T_0,\dots,T_3\}$ on $\cX\otimes\cY$ such that $T_i\ket{\Psi_j}=\delta_{ij}\ket{\Psi_j}$.

We begin with a simple observation. From Eq.~\eqref{eq:Wstates} it follows that
\begin{subequations}\label{eq:symm}
\begin{align}
\proj{\Psi_{k\oplus1}}&=W\proj{\Psi_k}W^\dagger\qquad(k=0,1,2),\\
\proj{\Psi_3}&=W\proj{\Psi_3}W^\dagger.
\end{align}
\end{subequations}
We take $T_3=\proj{\Psi_3}$, so that $T_3\ket{\Psi_j}=\delta_{3j}\ket{\Psi_j}$, and suppose that the remaining operators satisfy a relationship similar to Eq.~(\ref{eq:symm}a),\footnote{Measurements covariant under a local symmetry of the states were used for PPT discrimination in Refs.~\cite{YuDuanYing,Cosentino}.}
\begin{equation}\label{eq:cov}
T_{k\oplus1}=WT_kW^\dagger\qquad(k=0,1,2).
\end{equation}
This implies
\begin{equation}\label{eq:Tstates}
T_{k\oplus1}\ket{\Psi_{j\oplus1}}=WT_k\ket{\Psi_j},\qquad T_{k\oplus1}\ket{\Psi_3}=WT_k\ket{\Psi_3}\qquad(j,k=0,1,2).
\end{equation}
If $T_0\ket{\Psi_j}=\delta_{0j}\ket{\Psi_j}$, Eq.~\eqref{eq:Tstates} gives
\begin{subequations}
\begin{align}
T_1\ket{\Psi_{j\oplus1}}&=\delta_{0j}\ket{\Psi_{j\oplus1}}\qquad(j=0,1,2),\\
T_2\ket{\Psi_{j\oplus2}}&=\delta_{0j}\ket{\Psi_{j\oplus2}}\qquad(j=0,1,2),\\
T_1\ket{\Psi_3}&=T_2\ket{\Psi_3}=0,
\end{align}
\end{subequations}
which are the conditions for $T_1$ and $T_2$. Hence $\{T_k\}$ is a measurement that perfectly distinguishes the states if and only if $T_0$ satisfies
\begin{subequations}\label{eq:twirlboth}
\begin{align}
T_0\ket{\Psi_j}&=\delta_{0j}\ket{\Psi_j}\quad(j=0,\dots,3),\label{eq:detect}\\
T_0+WT_0W^\dagger+W^\dagger T_0W&=\id-\proj{\Psi_3} ,\label{eq:twirl}
\end{align}
\end{subequations}
where we arrive at Eq.~\eqref{eq:twirl} using Eq.~\eqref{eq:cov}, $T_3=\proj{\Psi_3}$, and the completeness condition $\sum_kT_k=\id$. We now look for a separable $T_0$ satisfying these equations.

\paragraph{Conditions on $T_0$.}
A separable $T_0$ can be written as
\begin{equation}\label{eq:T0sep}
T_0=\sum_i\tau_i\proj{u_iv_i},\qquad \tau_i>0 ,
\end{equation}
with product vectors $\ket{u_i}\ket{v_i}\in\cX\otimes\cY$. Since every term in the above sum is positive, Eq.~\eqref{eq:detect} for $j=1,2,3$ holds iff $\braket{\Psi_k}{u_i\,v_i}=0$ for $k=1,2,3$ and every $i$. To express this condition, let $\ket x\ket y\in\cX\otimes\cY$ be an arbitrary product vector, where
\begin{subequations}\label{eq:xy}
\begin{align}
\ket x&=a\ket{00}+b\ket{01}+c\ket{10}+d\ket{11}\in\cX,\\
\ket y&=e\ket{00}+f\ket{01}+g\ket{10}+h\ket{11}\in\cY .
\end{align}
\end{subequations}
Computing the overlap with the states~\eqref{eq:PsiXY} gives
\begin{subequations}\label{eq:overlap}
\begin{align}
\braket{\Psi_k}{x\,y}&=\tfrac13\left(F_0+\omega^{2k}F_1+\omega^kF_2\right)\qquad(k=0,1,2),\\
\braket{\Psi_3}{x\,y}&=dh,
\end{align}
\end{subequations}
where
\begin{equation}\label{eq:F}
F_0=ae+bg+cf,\qquad F_1=a(f+g)+de,\qquad F_2=ah+(b+c)e.
\end{equation}
For $\ket x\ket y=\ket{u_i}\ket{v_i}$, the conditions for $k=1,2$ are $F_0+\omega^2F_1+\omega F_2=0$ and $F_0+\omega F_1+\omega^2F_2=0$, which hold iff $F_0=F_1=F_2$ since $1+\omega+\omega^2=0$. The condition for $k=3$ is $dh=0$. Thus
\begin{equation}\label{eq:admissible}
F_0=F_1=F_2\quad(k=1,2),\qquad dh=0\quad(k=3),
\end{equation}
and then $\braket{\Psi_0}{u_i\,v_i}=F_0$. Vectors satisfying Eq.~\eqref{eq:admissible} with $F_0=0$ are orthogonal to all four states.

The remaining conditions on $T_0$ come from Eq.~\eqref{eq:twirl}. First observe that
\begin{equation}\label{eq:Wphase}
W\ket{a_1a_2}\ket{b_1b_2}=\omega^q\ket{a_1a_2}\ket{b_1b_2},\qquad q=2(a_1+a_2)+(b_1+b_2)\pmod3 ,
\end{equation}
where $\ket{a_1a_2}\ket{b_1b_2}$ are the computational basis vectors of $\cX\otimes\cY$. Thus the eigenspaces of $W$ are spanned by these vectors. For any two of them, $\ket i$ and $\ket j$, with $W\ket i=\omega^{q_i}\ket i$ and $W\ket j=\omega^{q_j}\ket j$,
\begin{subequations}
\begin{align}
\bra iWT_0W^\dagger\ket j&=\omega^{q_i-q_j}\bra iT_0\ket j,\\
\bra iW^\dagger T_0W\ket j&=\omega^{q_j-q_i}\bra iT_0\ket j ,
\end{align}
\end{subequations}
so that
\begin{equation}
\bra i\bigl(T_0+WT_0W^\dagger+W^\dagger T_0W\bigr)\ket j=\bigl(1+\omega^{q_i-q_j}+\omega^{q_j-q_i}\bigr)\bra iT_0\ket j .
\end{equation}
Using Eq.~\eqref{eq:twirl} and the above equation, we therefore have
\begin{equation}\label{eq:factor}
\bigl(1+\omega^{q_i-q_j}+\omega^{q_j-q_i}\bigr)\bra iT_0\ket j=\delta_{ij}-\braket{i}{\Psi_3}\braket{\Psi_3}{j} .
\end{equation}
If $q_i=q_j$, Eq.~\eqref{eq:factor} gives
\begin{equation}\label{eq:blocks}
\bra iT_0\ket j=\tfrac13\bigl(\delta_{ij}-\braket{i}{\Psi_3}\braket{\Psi_3}{j}\bigr).
\end{equation}
If $q_i\ne q_j$ both sides of Eq.~\eqref{eq:factor} vanish, so these elements of $T_0$ are not constrained by Eq.~\eqref{eq:twirl}. Thus $T_0$ must satisfy Eq.~\eqref{eq:admissible} for each of its product vectors and Eq.~\eqref{eq:blocks}. Eq.~\eqref{eq:detect} for $j=0$ then follows from Eqs.~\eqref{eq:twirl} and~\eqref{eq:admissible}.

\paragraph{Solution.}
We have thus reduced the problem to finding product vectors $\ket{u_i}\ket{v_i}$ satisfying Eq.~\eqref{eq:admissible} and weights $\tau_i>0$ such that $T_0$ satisfies Eq.~\eqref{eq:blocks}. A solution is given below in closed form.

Specifically, $T_0=P_0+C$, where $P_0=\sum_rw_r\Pi_r$ is formed from the vectors and weights of Table~\ref{tab:reps}, and $C$ is an operator orthogonal to all four states and is given by
\begin{equation}\label{eq:C}
C=\tfrac14\bigl(\proj{11}\otimes\proj{\Psi^+}+\proj{\Psi^+}\otimes\proj{11}\bigr)+\tfrac13\bigl(\proj{\xi}\otimes\proj{\Psi^-}+\proj{\Psi^-}\otimes\proj{\xi}\bigr),
\end{equation}
where $\ket\xi=\tfrac12\bigl(\ket{00}+\ket{01}+\ket{10}-\ket{11}\bigr)$. It can be checked directly that the product vectors of Table~\ref{tab:reps} and of the operator~\eqref{eq:C} satisfy Eq.~\eqref{eq:admissible}, and that $T_0$ satisfies Eq.~\eqref{eq:blocks}.

By Eq.~\eqref{eq:cov} and $T_3=\proj{\Psi_3}$, the remaining operators are
\begin{subequations}
\begin{align}
T_1&=P_1+WCW^\dagger,\\
T_2&=P_2+W^\dagger CW,\\
T_3&=P_3,
\end{align}
\end{subequations}
with $P_1=WP_0W^\dagger$, $P_2=W^\dagger P_0W$ and $P_3=\proj{\Psi_3}$. Since $C$ is orthogonal to all four states and $W$ permutes them, $WCW^\dagger$ and $W^\dagger CW$ are orthogonal to all four states as well. We combine them to define the operator
\begin{equation}\label{eq:P4sum}
P_4=C+WCW^\dagger+W^\dagger CW .
\end{equation}
\begin{table}[t]
\centering
\small
\renewcommand{\arraystretch}{1.3}
\begin{tabular}{l l c c}
\toprule
Element & $\ket x\ket y$ & $\ket x\in\cX_1\otimes\cX_2$ & $\ket y\in\cY_1\otimes\cY_2$\\
\midrule
$P_0=\sum_{r,i}w_r\proj{x_{ri}y_{ri}}$ & $\ket{x_{ri}}\ket{y_{ri}}$, $r=1$--$4$, $i=1,3$ & entangled & product\\
 & $\ket{x_{ri}}\ket{y_{ri}}$, $r=1$--$4$, $i=2,4$ & product & entangled\\
 & $\ket{x_{ri}}\ket{y_{ri}}$, $r=5$, $i=1$--$4$ & entangled & entangled\\
\midrule
$P_1=WP_0W^\dagger$ & $W\ket{x_{ri}}\ket{y_{ri}}$ & \multicolumn{2}{c}{same as for $P_0$}\\
$P_2=W^\dagger P_0W$ & $W^\dagger\ket{x_{ri}}\ket{y_{ri}}$ & \multicolumn{2}{c}{same as for $P_0$}\\
\midrule
$P_3=\proj{\Psi_3}$ & $\ket{11}\ket{11}$ & product & product\\
\midrule
$P_4$ & $\ket{11}\ket{\Psi^+}$, $\ket{00}\ket{\Psi^-}$, $\ket{11}\ket{\Psi^-}$ & product & entangled\\
 & $\ket{\Psi^+}\ket{11}$, $\ket{\Psi^-}\ket{00}$, $\ket{\Psi^-}\ket{11}$ & entangled & product\\
 & $\ket{\Psi^+}\ket{\Psi^-}$, $\ket{\Psi^-}\ket{\Psi^+}$ & entangled & entangled\\
\bottomrule
\end{tabular}
\caption{Product vectors $\ket x\ket y$ of the measurement of Proposition~\ref{prop:collective}, and whether $\ket x$ and $\ket y$ are entangled across the two copies. The vectors of $P_0$ are those of Table~\ref{tab:reps}, and those of $P_4$ follow from Eq.~\eqref{eq:P4final}.}
\label{tab:ent}
\end{table}%
Simplifying (Appendix~\ref{app:P4}), we get
\begin{equation}\label{eq:P4final}
P_4=\tfrac34\bigl(\proj{11}\otimes\proj{\Psi^+}+\proj{\Psi^+}\otimes\proj{11}\bigr)+\tfrac14\bigl(M\otimes\proj{\Psi^-}+\proj{\Psi^-}\otimes M\bigr),
\end{equation}
where $M$ is defined in~\eqref{eq:PiM}. We therefore have
\begin{equation}\label{eq:Pfinal}
\sum_{i=0}^4P_i=\sum_{k=0}^3T_k=\id,\qquad P_i\ket{\Psi_j}=T_i\ket{\Psi_j}=\delta_{ij}\ket{\Psi_j},\qquad P_4\ket{\Psi_j}=0\qquad(i,j=0,\dots,3).
\end{equation}
Thus $\{P_0,\dots,P_4\}$ is the desired separable measurement. This proves Proposition~\ref{prop:collective}.

\section{Discussion}\label{sec:discussion}
Our result shows that perfect discrimination of an orthogonal basis of $\mathbb C^2\otimes\mathbb C^2$ by separable measurements may require collective measurements on two copies of the unknown state. The basis~\eqref{eq:omega-basis} is an example for which adaptive SEP protocols fail but a collective SEP measurement distinguishes the states perfectly.

Table~\ref{tab:ent} shows how the measurement uses the two copies. For $r=1$--$4$, either $\ket{x_{ri}}$ or $\ket{y_{ri}}$ of every product vector $\ket{x_{ri}}\ket{y_{ri}}$ of $P_0$ is entangled across the copies, while for $r=5$ both $\ket{x_{ri}}$ and $\ket{y_{ri}}$ are. The same holds for $P_1=WP_0W^\dagger$ and $P_2=W^\dagger P_0W$ because $W$ is a product of unitaries on $\cX_1$, $\cX_2$, $\cY_1$, and $\cY_2$. For every product vector $\ket x\ket y$ of $P_4$, at least one of $\ket x$ and $\ket y$ is also entangled across the copies, although this outcome never occurs and therefore plays no role in discrimination. Only $P_3=\proj{\Psi_3}$, which detects the product state, is a product operator across all four qubits.

A natural question is whether this separable measurement can be implemented by LOCC. We do not know the answer, and we have not been able to make much progress on it. Thus whether the basis~\eqref{eq:omega-basis} can be perfectly distinguished by LOCC with two copies remains open, and so does the general question of whether two copies are always sufficient for perfect LOCC discrimination of orthogonal states in bipartite systems.

\section*{Acknowledgments}
The solution for $T_0$ in Sec.~\ref{sec:sep} was found with the help of Claude (Anthropic). It was verified by the author.

\FloatBarrier
\appendix
\section{Derivation of \texorpdfstring{$H$}{H}}\label{app:witness}
We wish to find a Hermitian operator $H$ of nonzero trace on $\mathbb C^2\otimes\mathbb C^2$ whose expectation value vanishes on every vector in \eqref{eq:prodvecs}. Consider an operator of the form
\begin{equation}
H=\begin{pmatrix}h_{00}&0&0&\eta\\0&h_{01}&0&0\\0&0&h_{10}&0\\\eta&0&0&h_{11}\end{pmatrix}
\end{equation}
in the basis $\ket{00},\ket{01},\ket{10},\ket{11}$, where $h_{00},h_{01},h_{10},h_{11}$ and $\eta$ are real. For an arbitrary vector $\ket v=\sum_{i,j}v_{ij}\ket{ij}$, we then have
\begin{equation}
\bra vH\ket v=h_{00}|v_{00}|^2+h_{01}|v_{01}|^2+h_{10}|v_{10}|^2+h_{11}|v_{11}|^2+2\eta\operatorname{Re}(v_{00}^*v_{11}).
\end{equation}

Requiring this expectation value to vanish for the vectors in \eqref{eq:prodvecs} gives
\begin{align}
\ket{11}:\qquad&h_{11}=0,\\
\ket0\bigl(\ket0+c\ket1\bigr):\qquad&h_{00}+h_{01}=0,\\
\bigl(\ket0+c\ket1\bigr)\ket0:\qquad&h_{00}+h_{10}=0,\\
\bigl(\ket0+a\ket1\bigr)\bigl(\ket0+b\ket1\bigr):\qquad&h_{00}+h_{01}+h_{10}+h_{11}+2\eta=0,
\end{align}
where $|a|=|b|=|c|=1$ and $ab=1$, as in \eqref{eq:prodvecs}.

These equations give
\begin{equation}
h_{01}=h_{10}=-h_{00},\qquad h_{11}=0,\qquad \eta=\frac{h_{00}}{2}.
\end{equation}
Consequently, $\operatorname{Tr}H=-h_{00}$, which is nonzero whenever $h_{00}\neq0$. This is the operator~\eqref{eq:H} for $\eta=1$.

\section{Evaluation of \texorpdfstring{$P_4$}{P4}}\label{app:P4}
By~\eqref{eq:W} and Eq.~\eqref{eq:Wphase},
\[
W\ket{11}\ket{\Psi^+}=\omega^2\ket{11}\ket{\Psi^+},\qquad W\ket{\Psi^+}\ket{11}=\omega\ket{\Psi^+}\ket{11},
\]
so that
\begin{align*}
\sum_{k=0}^{2}W^k\bigl(\proj{11}\otimes\proj{\Psi^+}\bigr)W^{-k}&=3\proj{11}\otimes\proj{\Psi^+},\\
\sum_{k=0}^{2}W^k\bigl(\proj{\Psi^+}\otimes\proj{11}\bigr)W^{-k}&=3\proj{\Psi^+}\otimes\proj{11}.
\end{align*}
For the second term of the operator~\eqref{eq:C},
\[
(U_A\otimes U_A)\ket\xi=\tfrac12\bigl(\ket{00}+\omega^2\ket{01}+\omega^2\ket{10}-\omega\ket{11}\bigr),\qquad (U_B\otimes U_B)\ket{\Psi^-}=\omega\ket{\Psi^-}.
\]
Hence
\begin{equation}
\sum_{k=0}^{2}W^k\bigl(\proj{\xi}\otimes\proj{\Psi^-}\bigr)W^{-k}=\tfrac34\,M\otimes\proj{\Psi^-}.
\end{equation}
The same holds with $\cX$ and $\cY$ interchanged. With the weights $\tfrac14$ and $\tfrac13$ in~\eqref{eq:C}, the sum~\eqref{eq:P4sum} gives Eq.~\eqref{eq:P4final}.

\end{document}